\documentclass[doublecol]{epl2} 

\title{Molecular Dynamics Computer Simulation of Crystal Growth and Melting in Al$_{50}$Ni$_{50}$}
\shorttitle{Crystal Growth and Melting in Al$_{50}$Ni$_{50}$}

\author{A. Kerrache\inst{1} \and J. Horbach\inst{1,2} \and K. Binder\inst{1}}
\shortauthor{A. Kerrache \etal}

\institute{                    
  \inst{1} Institut f\"ur Physik, Johannes Gutenberg--Universit\"at Mainz, Staudinger Weg 7, 
           55099 Mainz, Germany\\
  \inst{2} Institut f\"ur Materialphysik im Weltraum, Deutsches Zentrum f\"ur Luft-- und 
           Raumfahrt (DLR), 51170 K\"oln, Germany
}
\pacs{81.10.Aj}{Theory and models of crystal growth; physics of crystal growth, crystal morphology and orientation}
\pacs{64.70.Dv}{Solid-liquid transitions}
\pacs{02.70.Ns}{Molecular dynamics and particle methods}

\abstract{
The melting and crystallization of Al$_{50}$Ni$_{50}$ are studied by
means of molecular dynamics computer simulations, using a potential of
the embedded atom type to model the interactions between the particles. Systems
in a slab geometry are simulated where the B2 phase of AlNi in the middle
of an elongated simulation box is separated by two planar interfaces from
the liquid phase, thereby considering the (100) crystal orientation.
By determining the temperature dependence of the interface velocity,
an accurate estimate of the melting temperature is provided. The value
$k=0.0025$\,m/s/K for the kinetic growth coefficient is found. This
value is about two orders of magnitude smaller than that found in recent
simulation studies of one-component metals. The classical Wilson-Frenkel 
model is not able to describe the crystal growth kinetics on a quantitative
level. We argue that this is due to the neglect of diffusion processes
in the liquid-crystal interface.}
\begin{document}

\maketitle

\section{Introduction}

The classical model for crystallization from the melt is the one
proposed by Wilson \cite{wilson00} and Frenkel \cite{frenkel32}. It
considers crystal growth as an activated process, controled
by the mass transport in the liquid. However, various studies
using molecular dynamics (MD) computer simulation have shown
that the Wilson-Frenkel scenario is not applicable to a large
class of materials. Especially in pure metals, growth kinetics
is much faster than expected for an activated diffusion-limited mechanism
\cite{broughton86,burke88,briels97,tepper01,huitema99a,huitema99b,hoyt02,sun04a,xia07,jesson01,celestini02,amini06}.
In this case, rearrangements in the liquid structure are not required
to provide the formation of crystalline layers.  This may explain why
one-component metals are not glassforming systems in general. On the
other hand, binary metallic alloys are known as glassforming systems,
provided that heterogeneous nucleation can be avoided.  Indeed, these
systems exhibit in general a much slower growth kinetics than pure
metals. Recent studies have demonstrated that the MD simulation technique
is well-suited to elucidate the crystallization kinetics in binary alloys
\cite{teichler99,asta99,ramalingam02,becker07a,becker07b,becker07c,sibug02}.
But these studies show also that the growth kinetics in binary mixtures
is more complicated than in the one-component counterparts.  One of
the open questions is to what extent the Wilson-Frenkel
picture is valid for two-component metals.  This question is addressed
in the following.

In this work, the crystal growth kinetics of the binary alloy
Al$_{50}$Ni$_{50}$ is investigated by MD simulation. The experimental
melting temperature for this system is at 1920\,K where it exhibits
a first order phase transition from a liquid to an intermetallic B2
phase. Very recently, the crystal growth velocity for this transition
has been measured by Reutzel {\it et al.}~\cite{reutzel07} using an
electromagnetic levitation technique under reduced gravity conditions in
combination with a high-speed camera.  At an undercooling of about 60\,K,
growth velocities of the order of 0.1\,m/s were found.  This value is
about two orders of magnitude smaller than that found for pure metals
at comparable undercoolings, indicating that Al$_{50}$Ni$_{50}$ may be
the prototype of a system with a diffusion-limited growth mechanism.

The MD simulation allows for an accurate determination of the melting
temperature, kinetic growth coefficients, and transport coefficients such
as self- and interdiffusion constants. These information are required
to check the validity of the Wilson-Frenkel model of crystal growth.
As we shall see below the kinetic growth coefficient, as estimated by
our simulation for the (100) orientation of the crystal, is indeed much
smaller than that found for simple metals. Thereby, the growth velocities
are in good agreement with those measured in the aforementioned experiment
by Reutzel {\it et al.}~\cite{reutzel07}. However, we demonstrate that
the Wilson-Frenkel model is not able to describe the crystal growth
kinetics in Al$_{50}$Ni$_{50}$ on a quantitative level, at least for
plausible choices of the various free parameters appearing in the
theory. This indicates the need of microscopic theories on the various
aspects of crystal growth kinetics. By computing diffusion profiles for an
inhomogeneous crystal-liquid system at coexistence, we show explicitely
that such theories have to take into account diffusion processes in the
crystal-liquid interface region.


%
\section{Details of the simulation}
To investigate the crystallization of Al$_{50}$Ni$_{50}$
from the melt, we have done extensive molecular dynamics
computer simulations.  The interactions between the atoms were modelled
by a potential of the embedded atom type, proposed by Mishin {\it et
al.}~\cite{mishin02}.  Recent studies have shown that this potential
gives a realistic description of the diffusion dynamics in Al-Ni melts
\cite{das05,horbach07}.  The simulations were done at constant pressure
($p_{\rm ext}=0$).  For this, an algorithm proposed by Andersen was used,
setting the mass of the piston to 0.0027\,u \cite{andersen80}. Temperature
was kept constant by coupling the system at every 100 steps to a
stochastic heat bath. The equations of motion were integrated with the
velocity form of the Verlet algorithm \cite{allen} with a time step of 1\,fs.

\begin{figure}[t]
\onefigure[scale=1.25]{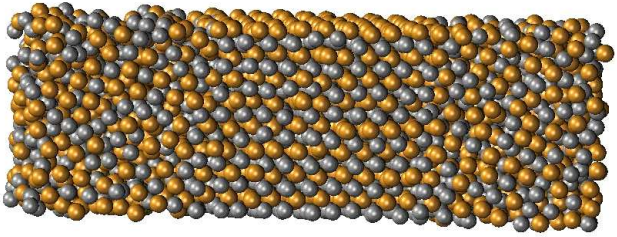}
\caption{\label{fig0} Snapshot of a simulated configuration with two
crystal-melt interfaces of the system Al$_{50}$Ni$_{50}$ at the
temperature $T=1500$\,K. Al and Ni atoms are shown as grey and brown
spheres, respectively.
}
\end{figure}
At each temperature in the range 1600\,K$\ge T \ge 1200$\,K, 12
independent samples with solid-liquid interfaces were prepared. To this
end, the B2 phase of Al$_{50}$Ni$_{50}$ was equilibrated at the target
temperature for 1\,ns. The simulations were done for a system of $N=3072$
particles ($N_{\rm Al}=N_{\rm Ni}=1536$) in an elongated simulation box
of size $L\times L\times L_z$ (with $L_z=3\times L$), considering the (100)
direction of the crystal.  Periodic boundary conditions were employed
in all three spatial directions.  Having relaxed the crystal sample,
one third of the particles in the middle of the box were fixed and the
rest of the system was melted during 500\,ps at $T=3000$\,K. Then, the
whole system was annealed at the target temperature for another 500\,ps,
before we started the production runs over 1\,ns in the $NpT$ ensemble.
A snapshot of the system with two interfaces at $T=1500$\,K is shown
in Fig.~\ref{fig0}. We did also 20 independent microcanonical runs of
the crystal-liquid system at the coexistence temperature $T=1520$\,K
(see below), starting from fully equilibrated samples. These runs,
each of them over 1\,ns, was used to study the diffusion dynamics in the
crystal-liquid interface region.

In addition, simulations of liquid samples were performed at the
temperatures $T=1200$\,K, 1300\,K, 1400\,K, 1500\,K, 1600\,K, 1800\,K,
and $2000$\,K, in order to determine self-diffusion coefficients as well
as the interdiffusion coefficient (see below).  In this case, systems of
2000 particles were placed in a cubic simulation box.  At each temperature,
equilibration runs over 1\,ns were done in the $NpT$ ensemble, followed
by microcanonical production runs over 23\,ns.

\section{Results}
\begin{figure}
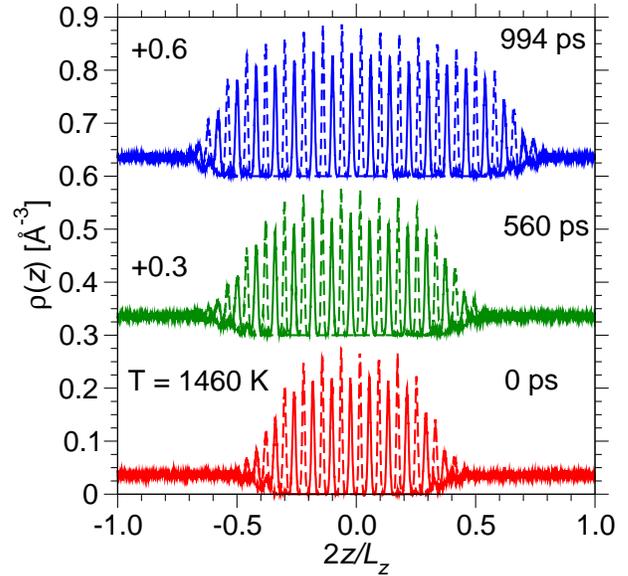

\onefigure[scale=0.42]{fig1.eps}
\caption{Number density profiles during crystal growth at $T=1460$\,K 
for Al (solid lines) and Ni (dashed lines). The profiles corresponding 
to $t=560$\,ps and $t=994$\,ps are shifted with respect to the 
$t=0$\,ps profiles by 0.3\,\AA~and 0.6\,\AA, as indicated.}
\label{fig1}
\end{figure}
As described in the previous section, samples in an elongated simulation
box were prepared as starting configurations where the crystal in the
middle is surrounded by the liquid phase on both sides, separated by two
interfaces (see Fig.~\ref{fig0}). The behavior of these samples depends
strongly on the temperature at which they are simulated. While below
the melting temperature $T_{\rm m}$, the crystal will grow (as shown in
Fig.~\ref{fig1}), it will melt above $T_{\rm m}$. From the simulation,
the velocity $v_{\rm I}$ with which the liquid-crystal interface moves
can be determined. At $T=T_{\rm m}$, the interface velocity $v_{\rm
I}$ vanishes. Thus, by the extrapolation $v_{\rm I}\to 0$ the melting
temperature $T_{\rm m}$ can be estimated. In the following, we show that
this procedure yields a rather accurate estimate of $T_{\rm m}$.  Then,
we demonstrate that the crystal growth mechanism in Al$_{50}$Ni$_{50}$ can
be elucidated by investigating the diffusion dynamics in the liquid phase
and in the crystal-liquid interface region.

Figure \ref{fig1} displays the partial number density
profiles $\rho(z)$ of Al and Ni at $T=1460$\,K along the $z$ direction, 
i.e.~perpendicular to the solid-liquid interfaces. The lower profiles in
Fig.~\ref{fig1} correspond to the starting configuration, while the
second and the third ones correspond to $t=560$\,ps and 994\,ps. Note that in
Fig.~\ref{fig1} the $z$ coordinate is scaled by the factor $2/L_z$, placing
$z=0$ in the middle of the simulation box. Whereas the crystal structure
leads to pronounced peaks in $\rho(z)$, a constant density is observed
for the liquid regions along the $z$ direction, as expected. We can also
infer from Fig.~\ref{fig1} that the intermetallic B2 phase [here in (100)
orientation] exhibits a pronounced chemical ordering, characterized
by the alternate sequence of Al and Ni layers.  This indicates that,
different from one-component metals, the crystal growth kinetics
relies on local rearrangements in the liquid structure. Thus, one may
expect that diffusive transport is required to bring the atoms of each
species to a suitable site in the B2 crystal.  As one can further see
in Fig.~\ref{fig1}, the crystal is growing at $T=1460$\,K. Thus, this
temperature is below the melting temperature of our Al$_{50}$Ni$_{50}$
model.

\begin{figure}
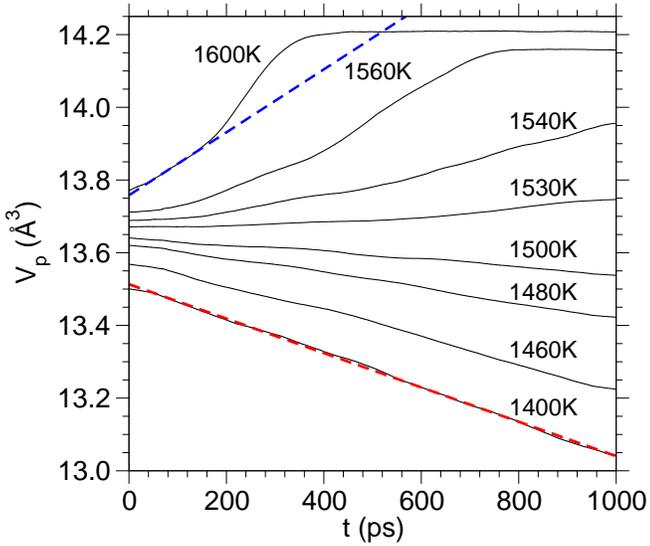

\onefigure[scale=0.4]{fig2.eps}
\caption{Volume per particle, $V_{\rm p}$ as a function of time for 
different temperatures, as indicated. The bold dashed lines are examples
of linear fits from which the volume velocity $\dot{V}$ is determined.}
\label{fig2}
\end{figure}
Since the density of the crystalline B2 phase is higher than that of the
liquid phase, the total volume of the system decreases at temperatures
$T<T_{\rm m}$ whereas it increases above $T_{\rm m}$. Figure \ref{fig2}
shows the time dependence of the volume per particle, $V_{\rm p}$,
for different temperatures between 1400\,K and 1600\,K. From this
plot, one can infer that the melting temperature is between 1500\,K and
1530\,K. Also shown in Fig.~\ref{fig2} are examples of linear fits of the
form $f(t)=A-\dot{V}_{\rm p}t$. Such linear growth laws are expected for
steady state growth \cite{huitema99a}. We use these fits to determine
the change of the volume $\dot{V}$ per unit time. The deviations from
the linear behavior at short times reveal that the growth (or melting)
of the crystal is not yet in a steady state regime \cite{huitema99a}. At
high temperatures, we see a complete melting of the crystal and thus
the volume $V_{\rm p}$ reaches a constant at long times corresponding to
the specific volume of the liquid phase. Prior to this, the melting of
the crystal is faster than in the linear steady-state regime.  In this
intermediate regime the crystal has shrunk to such small dimensions
that we see essentially the interaction between the two interfaces in
the simulation box and thus strong deviations from steady state growth
are observed.

\begin{figure}
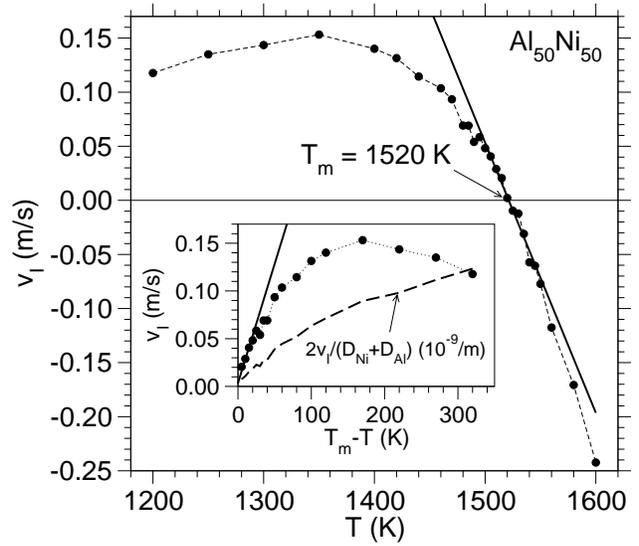

\onefigure[scale=0.4]{fig3.eps}
\caption{Interface velocity as a function of temperature (filled 
circles, the dashed line is a guide to the eye). The bold line is 
a linear fit, $v_{\rm I}=k (T_{\rm m}-T)$, yielding the kinetic
growth coefficient $k=0.0025$\,m/s/K and the melting temperature 
$T_{\rm m}=1520$\,K. The inset shows the interface velocity as a 
function of undercooling $T_{\rm m}-T$ and the interface velocity divided
by the averaged self-diffusion coefficient (dashed line).}
\label{fig3}
\end{figure}
From the volume change $\dot{V}_{\rm p}$, the velocity $v_{\rm I}$, with
which the liquid-crystal interfaces move, can be estimated as follows:   
\begin{equation}
  v_{\rm I}=\frac{\dot{V}_{\rm p}}{2 N_{\rm l} (V_{\rm c}-V_{\rm l})} d
\end{equation}
Here, the product $N_{\rm l} (V_{\rm c}-V_{\rm l})$ quantifies the increase of the
volume caused by the addition of a crystalline layer (with $N_{\rm l}$ the
average number of particles in a layer, and $V_{\rm c}$ and $V_{\rm l}$
the specific volumes of the crystal and the liquid phase, respectively).
The length $d$ is the spacing between crystalline layers.

Figure~\ref{fig3} displays the interface velocity $v_{\rm I}$ as a
function of temperature. We see that $v_{\rm I}$ vanishes around 1520\,K
and thus this temperature is the estimate for the melting temperature,
$T_{\rm m}$, of our simulation model. Note that the experimental value for
$T_{\rm m}$ is around 1920\,K and so our simulation underestimates the
experimental value by about 20\%. Around $T_{\rm m}$, the simulation
data for $v_{\rm I}$ can be fitted by the linear law $v_{\rm I}=k
(T_{\rm m}-T)$ where the fit parameter $k$ is the so-called kinetic
coefficient. The fit, that is shown in Fig.~\ref{fig3}, yields the value
$k=0.0025$\,m/s/K. This value is about two orders of magnitude smaller
than the typical values for kinetic coefficients that have been found
in simulations of one-component metals \cite{hoyt02,sun04a,xia07,jackson02}.

The inset in Fig.~\ref{fig3} shows the interface velocity as a function
of undercooling $\Delta T =T_{\rm m}-T$. We see that $v_{\rm I}$
increases linearly up to an undercooling of about 30\,K. At $\Delta T
\approx 180$\,K, the interface velocity reaches a maximum value of about
0.15\,m/s. Note that at small undercoolings our simulation data are in
good agreement with recent experimental data on Al$_{50}$Ni$_{\rm 50}$,
measured under reduced gravity conditions during a parabolic flight
campaign \cite{reutzel07}.  Also shown in the inset of Fig.~\ref{fig3}
is the quantity $2v_{\rm I}/(D_{\rm Ni}+D_{\rm Al})$, with $D_{\rm Ni}$
and $D_{\rm Al}$ the self-diffusion constants of Ni and Al, respectively.
The self-diffusion constants will be discussed in detail below. Here,
we note that the maximum in $v_{\rm I}$ disappears when one divides
this quantity by the averaged self-diffusion coefficient. Thus, the
occurrence of a maximum in $v_{\rm I}(\Delta T)$ is due to the slowing
down of diffusion processes with decreasing temperature.

On a qualitative level, the behavior of $v_{\rm I}(\Delta T)$ can be
understood in the framework of the Wilson-Frenkel model.  The model
relates the interface velocity to the difference between the rate at
which the atoms join the crystal and the rate at which they leave the
crystal.  As a result the following formula for $v_{\rm I}$ is obtained
\cite{jackson02}:
\begin{equation}
  v_{\rm I} = A_{\rm kin} \left[ 1- \exp\left(- \frac{\Delta g}{k_B T} 
              \right) \right]
\label{eq_wf}
\end{equation}
with $A_{\rm kin}$ a kinetic prefactor, $k_{\rm B}$ the Boltzmann
constant and $\Delta g$ the free energy difference between the liquid
and crystal phase. Close to coexistence, the free energy difference
$\Delta g$ is proportional to $\Delta T$, and the exponential function in
Eq.~(\ref{eq_wf}) can be approximated, such that $1- \exp(- \frac{\Delta
g}{k_B T})\approx \frac{l \Delta T}{k_B T T_{\rm m}}$ with $l$ the latent
heat of the liquid-to-solid transition. Furthermore, the kinetic prefactor
$A_{\rm kin}$ can be expressed in terms of the diffusion coefficient $D$
of the liquid. Eventually, at small $\Delta T$ the expression for $v_{\rm
I}$ can be written as \cite{jackson02}
\begin{equation}
   v_{\rm I}= k_{\rm WF} \Delta T \quad \quad {\rm with} \quad \quad
    k_{\rm WF}=\frac{6df}{\Lambda^2} D \frac{l}{k_BT T_{\rm m}}
 \label{eq_wf2}
\end{equation}
where $f$ represents the fraction of collisions with the crystal
that contribute to the growth of the crystal. The parameter $\Lambda$
corresponds to an elementary diffusive jump distance of particles in the
liquid \cite{jackson02}.  Note that it is assumed in the derivation of
Eq.~(\ref{eq_wf2}) that the diffusion constant can be expressed by an
Arrhenius law,
\begin{equation}
  D = D_0 \exp\left( - \frac{Q}{k_BT} \right)
 \quad \quad {\rm with} \quad \quad D_0 = \frac{1}{6} \Lambda^2 \nu
\label{eq_arr}
\end{equation}
with $Q$ an activation energy associated with the diffusion of the atoms in
the liquid and $\nu$ a frequency of the order of the Debye frequency.

\begin{figure}
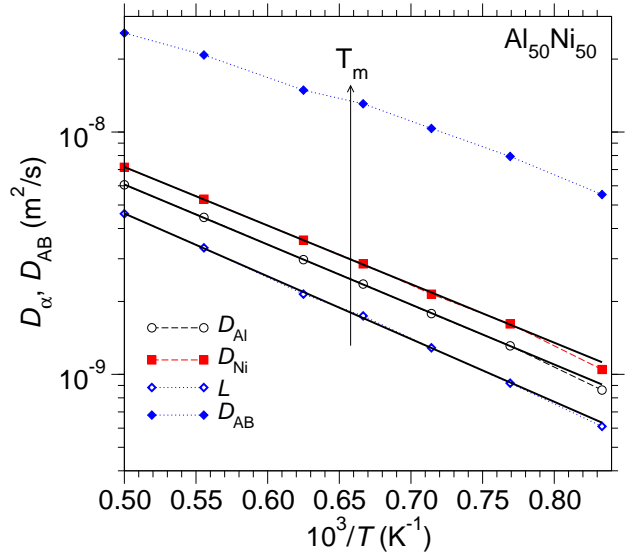

\onefigure[scale=0.4]{fig4.eps}
\caption{Arrhenius plot of self-diffusion constants $D_{\alpha}$ 
($\alpha={\rm Al, Ni}$), Onsager coefficient $L$ and interdiffusion 
constant $D_{\rm AB}$ for Al$_{50}$Ni$_{50}$. The solid lines are fits
with the Arrhenius law (\ref{eq_arr}), see text. The arrow indicates 
the location of the melting temperature $T_{\rm m}\approx1520$\,K 
of the simulation model.}
\label{fig4}
\end{figure}
In order to check whether the Wilson-Frenkel formula for the kinetic
coefficient $k_{\rm WF}$ in Eq.~(\ref{eq_wf2}) yields quantitative
agreement with the value $k=0.0025$\,m/s/K for Al$_{50}$Ni$_{50}$
(see above), we have computed the temperature dependence of self- and
interdiffusion coefficients around $T_{\rm m}=1520$\,K.  Whereas the
self-diffusion constant $D_{\alpha}$ is the transport coefficient
for tagged particle diffusion of atoms of type $\alpha$ (here
$\alpha={\rm Al, Ni}$), the interdiffusion coefficient $D_{\rm AB}$
describes diffusive transport due to concentration fluctuations among
the different components.  The self-diffusion constants $D_{\alpha}$
have been computed from the long-time limit of the corresponding
mean-squared displacements. The interdiffusion coefficient is given by
$D_{\rm AB}=\Phi L$ where $\Phi$ is the so-called thermodynamic factor
and $L$ is the Onsager coefficient. The thermodynamic factor expresses
the thermodynamic forces to homogenize the mixture with respect to
concentration fluctuations. We have calculated this quantity from the
$q\to 0$ limit of the inverse concentration-concentration structure factor
(see Ref.~\cite{horbach07}).  The Onsager coefficient $L$ contains all
the kinetic contributions to $D_{\rm AB}$ and can be determined from a
generalized mean-squared displacement describing the centre-of-mass motion of
one species. For details of the calculation of $L$ and $\Phi$, we refer
the reader to a recent publication \cite{horbach07}.

An Arrhenius plot of the different diffusion coefficients is shown in
Fig.~\ref{fig4}. As in a recent simulation study of Al$_{80}$Ni$_{20}$
\cite{horbach07}, the interdiffusion coefficient is about a factor
4 to 6 higher than the self-diffusion constants. This is due to the
thermodynamic factor (note that the Onsager coefficient lies below
the self-diffusion constants). The origin of this behavior is a large
resistance to macroscopic concentration fluctuations in dense liquids (a
similar property of dense liquids is their very low compressibility).
With respect to crystal growth kinetics in a binary alloy such as
Al$_{50}$Ni$_{50}$, it is not clear whether one has to consider self- or
interdiffusive transport as the limiting growth mechanism. 

Also shown in Fig.~\ref{fig4} are fits with Arrhenius laws (\ref{eq_arr}).
From these fits, we obtain the activation energies $Q=0.49$\,eV for
$D_{\rm Al}$, $Q=0.48$\,eV for $D_{\rm Ni}$, and $Q=0.51$\,eV for $L$.
The prefactors $D_0$ are $1.05\times 10^{-7}$\,m$^2$/s, $1.15\times
10^{-7}$\,m$^2$/s, and $0.91\times 10^{-7}$\,m$^2$/s for $D_{\rm Al}$,
$D_{\rm Ni}$, and $L$, respectively.  These values for the prefactors
can be compared to those proposed by Eq.~(\ref{eq_arr}).  With the
reasonable choices $\lambda=3$\,\AA~and $\nu=6$\,THz, similar values
for $D_0$ as in the fits are obtained, i.e.~$D_0\approx10^{-7}$.

\begin{figure}
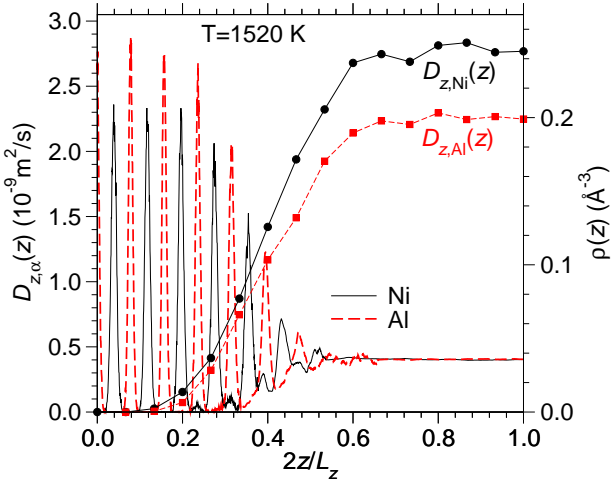

\onefigure[scale=0.35]{fig5.eps}
\caption{Number density profiles and diffusion profiles for Ni and Al
at coexistence, as indicated.}
\label{fig5}
\end{figure}
Moreover, the expression (\ref{eq_wf2}) for the kinetic coefficient
$k_{\rm WF}$ does not predict the order of magnitude correctly.
To see this, we can compute the value of $k_{\rm WF}$ at $T_{\rm
m}\approx1520$\,K using the results from the simulation. With
$l=0.23$\,eV, $D_{\alpha}\approx 2 \times 10^{-9}$\,m$^2$/s,
$\Lambda^2/6\approx 1.5$\,\AA$^2$, $d=3$\,\AA~and $f=1$, the value $k_{\rm
WF}\approx 0.05$\,m/s/K is yielded which is about one order of magnitude
higher than the value for $k$, as obtained in our simulation. The result
for $k_{\rm WF}$ is even worse if we replace the self-diffusion constant
by the interdiffusion constant in our estimate. In this case, we find
$k_{\rm WF}\approx 0.25$\,m/s/K.

But why does the Wilson-Frenkel theory overestimate the speed of crystal
growth?  To address this question we propose the following scenario: We
assume that the speed of crystal growth is limited by the atoms in the
liquid-crystal interface region and not by the atoms in the liquid region
where the liquid behaves like a bulk liquid. If this is true, one has to
study the diffusion dynamics in the interface region: if diffusion in the
interface region is much slower than in the bulk liquid, a failure of the
Wilson-Frenkel model would be plausible since this model only takes into
account diffusive transport of the bulk liquid.  To check this scenario,
we have simulated inhomogeneous systems with two crystal-liquid interfaces
at the melting temperature $T_{\rm m}=1520$\,K.  From these runs, we
determined the diffusion profiles $D_{z,\alpha}(z)$ ($\alpha={\rm Ni,
Al}$) along the $z$ direction that are shown in Fig.~\ref{fig5} together
with the number density profile.  $D_{z,\alpha}(z)$ was computed from
the long-time limit of the mean squared displacement in $z$ direction,
\begin{equation}
D_{z,\alpha}(z_s)= \lim_{t\to \infty} \frac{1}{N_{s}}
                  \sum_{i_{s}=1}^{N_{s}}
   \frac{\langle \left( z_{i_s}(t)-z_{i_s}(0)\right)^2 \rangle}{2t},
\end{equation}
where $z_{i_s}$ is the $z$ coordinate of a tagged particle that was
at time $t=0$ in one of 30 slabs that we introduced along the $z$
direction, each slab having a thickness of about 2.4\,\AA. $N_{s}$
is the number of particles in slab $s$ ($s=\{1, ..., 30\}$). As can be
seen in Fig.~\ref{fig5}, the interface region extends over 5-6 atomic
layers. Within this region the self-diffusion constants decrease roughly
by about one order of magnitude. When one considers crystal growth,
this slowing down of diffusion has to be taken into account, since the
formation of new crystalline layers occurs in the interface region.
This can be the reason why the Wilson-Frenkel model overestimates 
the speed of crystal growth.

\section{Conclusions}
Extensive MD simulations have been used to investigate the crystallization
kinetics as well as the diffusion dynamics of Al$_{50}$Ni$_{50}$.
Although crystal growth is relatively slow in this system, the simulation
yields accurate estimates of the melting temperature and the kinetic
growth coefficient [for the (100) orientation of the intermetallic B2
phase]. The small value of the latter quantity, $k=0.0025$\,m/s/K,
reveals that the growth kinetics of the intermetallic B2 phase is
controled by diffusive mass transport. However, the classical model
for diffusion-limited growth due to Wilson and Frenkel does not give
an accurate description. We argue that this is due to the neglect of
diffusive transport in the crystal-liquid interface region.  Microscopic
theories of crystal growth shall take into account the latter diffusive
processes.

\acknowledgments
Valuable discussions with Dieter Herlach and Andreas Meyer are gratefully
achnowledged.  We gratefully acknowledge financial support within the
Priority Program 1120 Phase Transformations in Multicomponent Melts of
the Deutsche Forschungsgemeinschaft (DFG).  Computing time on the JUMP
at the NIC J\"ulich and on the workstations at the ZDV (University of
Mainz) is gratefully acknowledged.

\end{document}